\documentclass[showpacs, amsfonts, amssymb, amsmath, 12pt, fleqn]{iopart}

\usepackage{iopams}
\usepackage{cite}
\usepackage{subfig}
\usepackage{float}
\usepackage{graphicx}						% Include figure files
\usepackage{epstopdf}
\usepackage{dcolumn}						% Align table columns on decimal point
\usepackage{bm}									% bold math
\usepackage[usenames]{color}		% Allow colored text using DVIPS names
\usepackage{multirow}
\usepackage[mathlines]{lineno}	% Enable numbering of text and display math
\usepackage{pifont}							% Enable PS ZapfDingbats

\newcommand{\lp}{\left(}
\newcommand{\rp}{\right)}
\newcommand{\lab}{\left<}
\newcommand{\rab}{\right>}
\newcommand{\lsb}{\left[}
\newcommand{\rsb}{\right]}

\newcommand{\labs}{\left|}
\newcommand{\rabs}{\right|}

\newcommand{\abs}[1]{\labs #1 \rabs}

\newcommand{\rn}{\rho_n}

\newcommand{\logten}{\log_{10}}

\newcommand{\dnene}{\delta n_{e}/n_{e}}
\newcommand{\dnbnb}{\delta n_{b}/n_{b}}

\newcommand{\dtete}{\delta T_{e}/T_{e}}

\newcommand{\dalpha}{D_{\alpha}}

\newcommand{\dne}{\delta n_{e}}
\newcommand{\dI}{\delta I}

\newcommand{\vti}{\ensuremath{v_{{\rm th}i}}}

\newcommand{\rhoi}{\rho_i}

\newcommand{\rhost}{\rho_{*}}

\newcommand{\LTi}{L_{T_i}}
\newcommand{\Lne}{L_{n_e}}

\newcommand{\RLTi}{R/\LTi}

\newcommand{\omgphi}{\omega_\phi}
\newcommand{\Uphi}{U_\phi}
\newcommand{\Uphipr}{U_\phi^\prime}
\newcommand{\Uperp}{U_\perp}

\newcommand{\Ubes}{U_{Z}^{\rm{BES}}}
\newcommand{\Uapp}{U_{Z}^{\rm{app}}}

\newcommand{\Udiai}{U^{\rm dia, \it i}}

\newcommand{\UphiC}{\Uphi^{\lp C\rp}}
\newcommand{\UphiD}{\Uphi^{\lp i\rp}}

\newcommand{\UappC}{U_{Z}^{\rm app, \it \lp C\rp}}

\newcommand{\Bphi}{B_\phi}
\newcommand{\Btheta}{B_\theta}

\newcommand{\gE}{\gamma_E}
\newcommand{\gEbar}{\bar\gamma_E}

\newcommand{\gm}{\gamma_{max}}

\newcommand{\lx}{\ell_x}
\newcommand{\ly}{\ell_y}
\newcommand{\lxy}{\ell_{x, y}}
\newcommand{\lZ}{\ell_Z}

\newcommand{\lpar}{\ell_\parallel}

\newcommand{\dlbeam}{\Delta L_b}

\newcommand{\tc}{\tau_{\rm c}}
\newcommand{\tstar}{\tau_*}

\newcommand{\tst}{\tau_{\rm st}}
\newcommand{\tM}{\tau_{\rm M}}
\newcommand{\tsh}{\tau_{\rm sh}}
\newcommand{\taupeak}{\Dt_{\rm peak}}
\newcommand{\tnl}{\tau_{\rm nl}}
\newcommand{\tnlnz}{\tau_{\rm nl}^{\rm NZ}}

\newcommand{\omgsti}{\omega_{\star i}}

\newcommand{\nuii}{\nu_{ii}}

\newcommand{\phizf}{\varphi^{\rm ZF}}
\newcommand{\phidw}{\varphi^{\rm DW}}

\newcommand{\UZ}{U_{Z}}
\newcommand{\ExB}{\bm{E} \times \bm{B}}
\newcommand{\UExB}{U^{\ExB}}
\newcommand{\UZExB}{U_Z^{\ExB}}

\newcommand{\kper}{k_\perp}

\newcommand{\kpri}{\kper \rhoi}
\newcommand{\kx}{k_x}
\newcommand{\ky}{k_y}

\newcommand{\corr}{\mathcal{C}}
\newcommand{\acov}{\mathcal{A}}
\newcommand{\dlnidlnn}{\mathcal{B}}
\newcommand{\order}{\mathcal{O}}

\newcommand{\Dx}{\Delta x}

\newcommand{\DZ}{\Delta Z}
\newcommand{\DR}{\Delta R}
\newcommand{\Dt}{\Delta t}

\newcommand{\psin}{\psi_N}
\newcommand{\psinh}{\psin^{1/2}}

\newcommand{\Qiturb}{\tilde{Q_i}}
\newcommand{\Qiexp}{Q_i^{exp}}
\newcommand{\Qisim}{Q_i^{sim}}
\newcommand{\QiNC}{Q_i^{NC}}
\newcommand{\QiGB}{Q_i^{GB}}

\newcommand{\fBW}{f_{\rm{BW}}}
\newcommand{\GAPD}{G_{\rm{APD}}}
\newcommand{\Imamp}{\Im_{\rm{amp}}}
\newcommand{\Qeff}{Q_{\rm{eff}}}
\newcommand{\FT}{F_{T}}
\newcommand{\FN}{F_{N}}

\newcommand{\DIIID}{\mbox{DIII-D}}

\newcommand{\figref}[1]{Fig.~\ref{fig:#1}}
\newcommand{\eqref}[1]{Eq.~(\ref{eq:#1})}
\newcommand{\refref}[1]{Ref.~\cite{#1}}
\newcommand{\tabref}[1]{Table~\ref{table:#1}}
\newcommand{\secref}[1]{\S\ref{sec:#1}}

\newcommand{\quant}[2]{#1\,\rm{#2}}

\begin{document}

\title[Comparison of turbulence measurements with direct gyrokinetic simulations]{Comparison of BES measurements of ion-scale turbulence with direct gyrokinetic simulations of MAST L-mode plasmas}

\author{A R Field$^1$, D Dunai$^2$, Y-c Ghim$^{1, 3}$, P Hill$^5$, B McMillan$^5$, \newline C M Roach$^1$, S Saarelma$^1$, A A Schekochihin$^4$, S Zoletnik$^2$ \newline and the MAST team}

\address{$^1$EURATOM/CCFE Fusion Association, Culham Science Centre, Abingdon, Oxon, OX14 3DB, UK.}

\address{$^2$Wigner Research Centre for Physics, HAS, H-1525, Budapest, Hungary}

\address{$^3$Korea Advanced Institute of Science and Technology (KAIST) Nuclear and Quantum Engineering, Daejeon, Republic of Korea, 305-701}

\address{$^4$Rudolf Peierls Centre for Theoretical Physics, University of Oxford, Oxford, UK.}

\address{$^5$Centre for Fusion, Space and Astrophysics, Warwick University, Coventry, UK}

\ead{anthony.field@ccfe.ac.uk}

\date{Oxford, \today}
 
\begin{abstract}
Observations of ion-scale ($\ky\rhoi \leq 1 $) density turbulence of relative amplitude $\gtrsim 0.2\%$ are available on the Mega Amp Spherical Tokamak (MAST) using a 2D (8 radial $\times$ 4 poloidal channel) imaging Beam Emission Spectroscopy (BES) diagnostic. Spatial and temporal characteristics of this turbulence, i.e., amplitudes, correlation times, radial and perpendicular correlation lengths and apparent phase velocities of the density contours, are determined by means of correlation analysis. For a low-density, L-mode discharge with strong equilibrium flow shear exhibiting an internal transport barrier (ITB) in the ion channel, the observed turbulence characteristics are compared with synthetic density turbulence data generated from global, non-linear, gyro-kinetic simulations using the particle-in-cell (PIC) code NEMORB. This validation exercise highlights the need to include increasingly sophisticated physics, e.g., kinetic treatment of trapped electrons, equilibrium flow shear and collisions, to reproduce most of the characteristics of the observed turbulence. Even so, significant discrepancies remain: an underprediction by the simulations of the turbulence amplituide and heat flux at plasma periphery and the finding that the correlation times of the numerically simulated turbulence are typically two orders of magnitude longer than those measured in MAST. Comparison of these correlation times with various linear timescales suggests that, while the measured turbulence is strong and may be `critically balanced', the simulated turbulence is weak.
\end{abstract}


\submitto{\PPCF}

% Comment out if separate title page not required
\maketitle

\section{Introduction}

A `grand challenge' of fusion research is to develop reliable, first-principles predictive capability of plasma confinement, e.g., of the ITER device. An essential part of this process is to perform quantitative comparisons of simulation results with experimental observations and thereby assess their validity \cite{tang_pop_2002}. Prediction of heat and particle fluxes require non-linear simulations of the saturated state of plasma turbulence, which can be validated at one level by comparing with the results of transport analysis and, at a deeper level, with measured turbulence characteristics. An excellent overview of the methodology of such studies can be found in \refref{rhodes_nf_2011}.

The micro-instabilities responsible for anomalous transport exist over a wide range of spatial scales: from electro-static ion-temperature-gradient (ITG), trapped-electron (TEM) and parallel-velocity gradient (PVG) modes (or electro-magnetic micro-tearing modes at high-$\beta$) at scales larger than the ion Larmor radius $\ky \rhoi \leq 1$ (where $\ky$ is the wavenumber perpendicular to the magnetic field, within a flux surface and $\rhoi$ is the ion Larmor radius), down to electron-temperature-gradient (ETG) modes with $\ky \rhoi \gg 1$ \cite{connor_ppcf_1994}. A wide range of diagnostic techniques is required to detect the resulting fluctuations in plasma parameters. Such verification programmes have been the focus of much research activity on conventional tokamaks, e.g., on \DIIID, which has a comprehensive set of turbulence diagnostics \cite{white_pop_2010, rhodes_nf_2011, shafer_pop_2012} and on Tore Supra \cite{casati_prl_2009, bourdelle_nf_2011}. Studies on the NSTX spherical torus have focused on electron-scale turbulence, comparing data from a micro-wave scattering system with global, non-linear gyro-kinetic simulations using GYRO \cite{ren_nf_2012} (earlier studies focussed on linear gyro-kinetic calculations \cite{mazzucato_prl_2008, yuh_prl_2011}). The availability of ion-scale density fluctuation measurements from BES \cite{smith_rsi_2010} will facilitate future multi-scale validation studies on the NSTX device. \footnote{See for example \refref{smith_pop_2013}, where study of inter-ELM turbulence in the pedestal region of H-mode plasmas in NSTX was carried out using this system and the parametric dependences of the spatial and temporal characeristics were compared with expectations for different types of turbulence.}

On DIII-D, a unique array of multi-scale, multi-field turbulence diagnostics has facilitated detailed validation studies comparing gyro-kinetic turbulence simulations with experimental data from L- and H-mode plasmas \cite{holland_jop_2008, holland_pop_2009, holland_pop_2011}. A variety of studies have been undertaken, e.g., of the dependence of the turbulence characteristics on $T_e/T_i$ \cite{rhodes_nf_2011} and elongation \cite{petty_pop_2008} and of the cross-phase between $\delta n_e$ and $\delta T_e$ fluctuations \cite{white_pop_2008}, thus probing the underlying physics, especially in cases where disagreement was found. Meaningful comparison of simulation results with experiment is only possible by implementing synthetic diagnostics that mimic the instrumental characteristics of the various fluctuation measurements, as was done, e.g., on DIII-D for BES and Correlation Electron Cyclotron Emission (CECE) diagnostics for measurements of $\dnene$ and $\dtete$, respectively \cite{shafer_rsi_2006, white_pop_2008, holland_pop_2009}. In the L-mode studies, although good agreement was found between simulated and measured heat flux and fluctuation characteristics in the mid-core region ($0.4 < r/a < 0.75$, where $r/a$ denotes the normalised radius), simulations generally underpredict the heat fluxes and amplitudes by almost an order of magnitude in the peripheral region ($r/a > 0.75$) \cite{white_pop_2008, holland_pop_2009, white_pop_2010, holland_pop_2011, rhodes_nf_2011}. This discrepancy has been attributed to increasing importance of edge-core coupling \cite{lin_pop_2004} or to an inward propagation of turbulence (or `avalanche') from the plasma edge \cite{diamond_nf_2001}. More recently, there have been concerted efforts to resolve this discrepancy, which now appears not to be ubiquitous.

No such validation excercise using global, non-linear simulations of ion-scale turbulence has as yet been performed for a spherical-tokamak plasma, whose particular characteristics make this especially interesting. Heating of the low-aspect-ratio plasma with tangentially directed neutral-beam-injection (NBI) heating results in strong equilibrium rotation in the core (toroidal Mach number $M_\phi = R\omgphi/\vti \lesssim 0.5$, where $\omgphi$ is the toroidal rotation rate and $\vti$ is the ion thermal velocity) and hence strong $\ExB$ shear $\gE = (\epsilon/q)\,d(R \omgphi)/dr$, where $\epsilon = r/R$ and $q$ the safety factor, which can be strong enough to stabilise ion-scale turbulence. The low aspect ratio also results in a large trapped particle fraction ($f_t \sim \epsilon^{1/2}$), enhancing the drive for Trapped-Electron Mode (TEM) micro-instabilities, although, as will be seen, collisions can reduce the trapped particle drive substantially in MAST plasmas. Finally, the low toroidal magnetic field $\Bphi$ results in larger values of the ion Larmor radius normalised to the plasma radius, $\rhost \equiv \rhoi/a$. The assumption of constant gradient scale lengths in local gyro-kinetic simulations is questionable when gradients vary appreciably over the simulation domain, which typically has to be several $10 \times \rhoi$ in radial extent to ensure convergence. Therefore, as discussed in \refref{saarelma_ppcf_2012}, for current-generation ST plasmas in which $\rhost \sim 1/50$, global codes may be required to perform meaningful non-linear simulations of the ion-scale turbulence.

Here we present the first validation exercise for a MAST plasma comparing the characteristics of ion-scale turbulence measured using a 2D BES turbulence imaging system \cite{field_rsi_2012} with synthetic data produced from non-linear simulations performed using the global particle-in-cell (PIC) code NEMORB \cite{joliet_cpc_2007}. The comparison is performed for a low-density, L-mode plasma exhibiting an internal transport barrier (ITB), similar to those used for earlier studies of ITB formation and dynamics \cite{field_iaea_2004, field_nf_2011}, for which linear gyro-kinetic stability calculations have been performed using the local code GS2 \cite{kotschenreuther_cpc_1995}. This equilibrium configuration was also used as the basis for the earlier global, non-linear simulations presented in \refref{saarelma_ppcf_2012}. An L-mode plasma is ideal for these studies because, in the peripheral region, the equilibrium flow shear is too weak to stabilise the ITG turbulence fully.

The MAST device is a medium-sized, low-aspect-ratio tokamak (aspect ratio $A = R/a \sim 1.3$, plasma current $I_p \leq 1.2\,\rm{MA}$, toroidal field $\Bphi \leq 0.58\,\rm{T}$ at 0.7 m), which is equipped with tangentially directed NBI heating (injected power $P_{NB} \leq 3.8\, \rm{MW}$ at $\sim \quant{70}{keV}$ $D^0$ injection energy). Studies of transport in MAST are facilitated by the availability of high-resolution kinetic profile data from a suite of advanced diagnostics (see \secref{exp_obs}) and an integrated analysis chain ($MC^3$) to prepare this data for transport analysis using TRANSP \cite{hawryluk_cec_1980}. The BES turbulence imaging system on MAST has sufficient signal-to-noise ratio (SNR) to detect density fluctuations with relative magnitude $\dnene \gtrsim 0.2\%$ if appropriate correlation analysis is used. Calculations of the three-dimensional spatial response and sensitivity of the BES system, accounting for relevant physical effects were performed in \refref{ghim_rsi_2010}
as described in \secref{bes_system}. This is quantified in terms of two-dimensional point-spread functions (PSFs), which are used here to generate synthetic BES data from the gyro-kinetic simulation data. Previous results of an analysis of the BES data \cite{field_iaea_2012} to elicit the structure and dynamics of the ion-scale turbulence \cite{ghim_prl_2013} and the dependence of the ion temperature gradient on the magnetic configuration, rotational shear and heat flux \cite{ghim_nfl_2013} are pertinent to our discussion of the results presented here.

The remainder of the paper in structured as follows. The results of gyro-kinetic simulations of MAST L-mode equilibria are presented in \secref{gk_sim}, with a summary of results of previous linear calculations in \secref{lin_gk_sim} and the non-linear simulations in \secref{nonl_gk_sim}. Our experimental observations of the ion-scale turbulence are explained in \secref{exp_obs}, with a brief description of the capabilities of the BES system in \secref{bes_system} and an explanation of the correlation analysis used to determine the turbulence characteristics in \secref{corr_anal}. The method used for the generation of synthetic data is then explained in \secref{synth_data}. The observed and simulated turbulence characteristics are then compared in \secref{nl_comp}. Finally, our conclusions are presented in \secref{conclusions}.

\section{\label{sec:gk_sim} Gyro-kinetic simulations of an L-mode discharge}

The validation study presented here is for an L-mode discharge formed using a scenario with early NBI heating during the current ramp, which slows current penetration, giving rise to strong toroidal rotation and negative magnetic shear ($\hat{s} = (r/q)(dq/dr)$) in the plasma core. Previous linear-stability calculations for such a discharge \cite{field_nf_2011} revealed the peripheral region, where the flow shear is much weaker, to be unstable to ITG modes. Therefore, such discharges hosting ITG turbulence are ideal candidates on which to validate non-linear, ion-scale turbulence simulations.

\begin{figure}[t]
\includegraphics[bb=0 0 440 146]{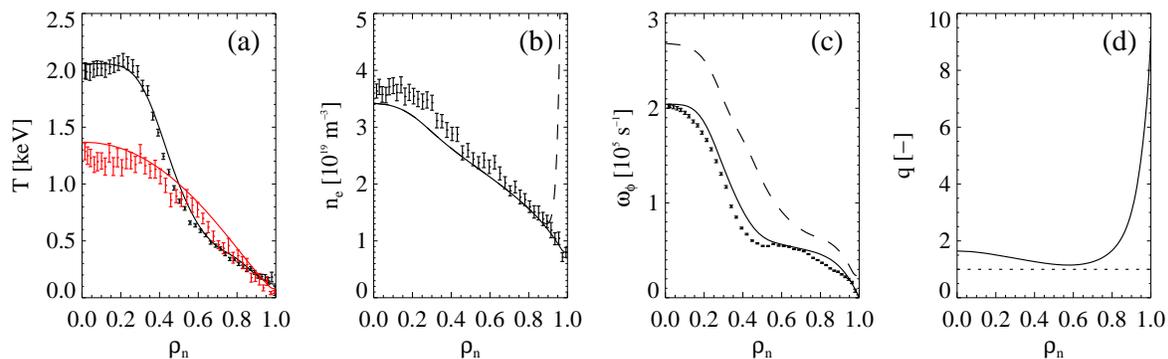}
\caption{\label{fig:27268_profiles} Equilibrium profiles for discharge \#27268 at \quant{0.25}{s}: (a) ion $T_i$ (black) and electron $T_e$ (red) temperatures, (b) electron density $n_e$, with the edge-shielding density (dashed, see \secref{numerics}), (c) toroidal rotation rate $\omgphi$ for the $C^{6+}$ ions (solid) and $D^+$ ions (dashed) and (d) the $q$-profile from EFIT. The raw data from the CXRS and TS systems is superposed where appropriate.}
\end{figure}

Kinetic profiles of the equilibrium used for this study are shown in \figref{27268_profiles}. They are taken from MAST L-mode discharge \#27268 at 0.25\;s, the parameters of which are: plasma current $I_p = \quant{800}{kA}$, toroidal field $\Bphi \sim \quant{0.58}{T}$ at $R_m = \quant{0.7}{m}$, co-injected NBI heating of $P_{NB} = \quant{3.3}{MW}$, line-average density $\bar{n_e} \leq \quant{2.3\times10^{19}}{m^{-3}}$, with $T_i \leq \quant{2.2}{keV}$, $T_e \leq \quant{1.5}{keV}$ and $\omgphi \leq \quant{2.0\times10^5}{rad/s}$, corresponding to Mach number $M_{\phi} \leq 0.5$. The foot of the ITB in the ion thermal and momentum channels is located at the normalised radius $\rn \equiv \psinh \sim 0.5$ (where $\psin$ is the normalised poloidal flux) in the region where $\hat{s} \leq 0$. Note that the temperature and rotation profiles of the bulk $D^+$ ions are calculated from measurements on the $C^{6+}$ impurities using the NCLASS \cite{houlberg_pop_1997} neo-classical package within the TRANSP code.

Linear-stability calculations were performed using both GS2 \cite{field_nf_2011} and NEMORB \cite{saarelma_ppcf_2012} for a similar equilibrium from an earlier MAST discharge (\#22087 at 0.25\;s), albeit with somewhat higher plasma current of $\quant{0.88}{MA}$, which was used for a study of ITB formation and evolution \cite{field_nf_2011}. The kinetic profiles from this equilibrium are very similar to those of the equilibrium studied here, so the results of these linear calculations are relevant and  briefly summarised in \secref{lin_gk_sim}. This equilibrium was also used as the basis for first global, non-linear simulations \cite{saarelma_ppcf_2012} using NEMORB and some of the relevant results are also summarised in \secref{nonl_gk_sim}.

\subsection{\label{sec:lin_gk_sim}Linear stability calculations of \#22807 at 0.25 s.}

As presented in \refref{field_nf_2011}, local linear-stability calculations with kinetic electrons were performed using GS2 for the equilibrium from discharge \#22807 at the time of peak ITB strength at $\quant{0.25}{s}$ at three locations: inside the ITB, just outside $q_{min}$ and in the plasma periphery. At the innermost of these surfaces (at $\Phi_N^{1/2}$ = 0.3, where $\Phi_N$ is the normalised toroidal flux, corresponding to $\rn \sim 0.36$) all modes at electron and ion scales were found to be stable in both electrostatic and electromagnetic calculations. At mid-radius ($\Phi_N^{1/2} = 0.52$, corresponding to $\rn = 0.67$), in the absence of flow shear, the relatively low collisionality results in appreciable trapped-electron drive and TEM modes are unstable in the intermediate wave-number range ($1 < \kpri < 10$) -- these are not completely stabilised by the flow shear. (In contrast, in calculations with adiabatic electrons, ITG modes that are unstable without flow shear are completely stabilised when the flow shear is included). At the outermost surface ($\Phi_N^{1/2} = 0.7$, corresponding to $\rn = 0.87$), the TEM modes are stable because the plasma is more collisional, while strongly growing ITG modes are not stabilised by the flow shear, which is weak at this radius.

To assess the importance of non-local effects, the results of linear stability calculations with NEMORB were compared with those from local, linear GS2 calculations (with and without collisions) \cite{saarelma_ppcf_2012}. Linear, electrostatic calculations were made with NEMORB (for the equilibrium from \#22807 at 0.25\;s) with adiabatic and kinetic electrons, both with and without flow shear \cite{mcmillan_pop_2011}, but without collisions. As in the GS2 calculations, the core was found to be stable to ITG modes, while the most unstable region was found to be $0.6 < \rn < 0.8$. By varying the normalised ion Larmor radius $\rhost$ in the NEMORB simulations, it was found that global effects significantly reduce the growth rates, starting at $\rhost$ just below the experimental value of 0.018.

\subsection{\label{sec:nonl_gk_sim}Global, non-linear simulations with NEMORB}

\subsubsection{\label{sec:nonl_gk_sim_22807}Previous simulations of \#22807 at 0.25 s.}

The results of global, electrostatic, non-linear simulations with NEMORB of the equilibrium from discharge \#22807 at $\quant{0.25}{s}$ were presented in \refref{saarelma_ppcf_2012} and the reader is referred to this article for details. In the simulations with adiabatic electrons but without sheared flow, the turbulence was found to spread from the linearly unstable region at $\rn \sim 0.7$ into the stable region down to $\rn \leq 0.4$ during the non-linear phase. The ion heat flux was found to be very low $\Qisim \leq \quant{0.01}{MW m^{-2}}$, which is below the neo-classical level. Increasing the normalised, inverse ion-temperature-gradient scale length $\RLTi = R\nabla T_i/T_i$ by 30\% increased the heat flux considerably, indicating that the gradient is close to the non-linear threshold for ITG turbulence. In simulations with kinetic electrons but without sheared flow, the heat flux increased to $\Qisim \leq \quant{1.2}{MW m^{-2}} $, which is far above the experimental level, $\Qiexp \sim \quant{0.02}{MW m^{-2}} $. Again, the turbulence spread into the linearly stable core region. Including equilibrium flow suppressed the turbulence in the region with strong shear ($\rn \leq 0.45$) and modestly reduced the peak heat flux. Results of runs incorporating electron collisions, which had been shown to reduce the linear drive due to trapped electrons \cite{roach_ppcf_2009}, were not reported in \refref{saarelma_ppcf_2012} (simulations with collisions have however been peformed for the more recent equilibrium discussed in \secref{nonl_gk_sim_27268}).

\subsubsection{\label{sec:nonl_gk_sim_27268}Non-linear simulations of \#27268 at 0.25 s.}

\begin{figure}[t]
\begin{center}
\includegraphics[width=3.5in]{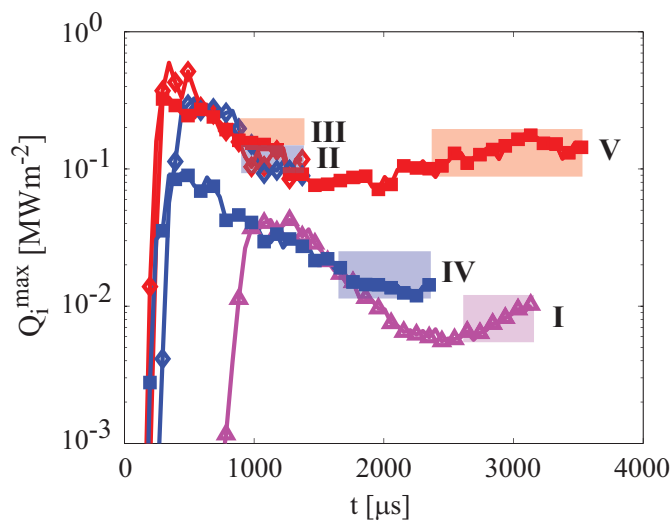}
\caption{\label{fig:qi_evo} The evolution of the maximum heat flux $Q_i^{max}$ during each of the non-linear NEMORB simulations listed in \tabref{sim_cases} where the plot symbols for the five cases are defined. The horizontal bars indicate the time periods used for the synthetic data generation.}
\end{center}
\end{figure}

In order to be able to compare our BES measurements with the non-linear simulations, further simulations have been carried out for an equilibrium from which the measurements are available, i.e., that corresponding to the kinetic profiles shown in \figref{27268_profiles}. Electrostatic simulations are available for five cases, distinguished by whether they were run with adiabatic (AE) or kinetic electrons (KE), with or without ion-electron and electron-electron collisions and with or without equilibrium toroidal flow. The various combinations for the five runs (\rm{I-V}) are summarised in \tabref{sim_cases}, where the start time and duration of the non-linear phase used to generate the synthetic data are also stated. As discussed in \secref{corr_anal} below, these periods are considered the minimum required to extract reasonable estimates of the turbulence characteristics. The evolution of the maximum heat flux $Q_i^{max}$ over the radial profile during each of the non-linear runs is shown in \figref{qi_evo}. 

\begin{table}[b]
\begin{center}
\begin{tabular}{c||c|c|c|c|c|c}
Case&KE&Flow&Coll.&Start [\rm{$\mu s$}]&Duration [\rm{$\mu s$}]&Symbol\\\hline
\rm{I}&$-$&$-$&$-$&2700&540&\textcolor{magenta}{$\triangle$}\\
\rm{II}&\ding{51}&$-$&$-$&850&489&\textcolor{blue}{$\diamond$}\\
\rm{III}&\ding{51}&\ding{51}&$-$&850&510&\textcolor{red}{$\diamond$}\\
\rm{IV}&\ding{51}&$-$&\ding{51}&1400&740&\textcolor{blue}{$\blacksquare$}\\
\rm{V}&\ding{51}&\ding{51}&\ding{51}&2350&1050&\textcolor{red}{$\blacksquare$}\\\hline
\end{tabular}
\caption{Summary of NEMORB simulation cases, showing the symbols used in Figs. 2-8 below. \label{table:sim_cases}}
\end{center}
\end{table}

\subsubsection{\label{sec:numerics} Details of numerical set-up.}

In NEMORB, the heat source is adjusted until the kinetic profiles match those prescribed and the resulting heat flux is a prediction which can be compared with the experimental value, although, as is discussed in \secref{comp_amp_qi}, a precise match cannot be expected. In order to maintain the signal-to-noise ratio $\geq 10$ for the duration of the simulation, $2 \times 10^8$ markers were used on an $N_x \times N_y \times N_\phi = 100 \times 512 \times 256$ grid, for the simulations with collisions the number of markers was doubled. For the simulations with kinetic electrons, the trapped-electron response is treated kinetically while that of the passing electrons is treated adiabatically. The reader is referred to \refref{saarelma_ppcf_2012} for further numerical details.

In order to ensure numerical stability, a boundary condition is adopted in NEMORB which forces both the density and potential perturbations at $\rn = 1$ to be zero, $\dnene = \varphi = 0$. Markers near the plasma boundary carry some density which is zeroed just before the Poisson equation is solved. Moreover, if a marker or one of its associated gyro-points lies outside the plasma, it is not taken into account. Therefore, the quasi-neutrality condition is violated near the boundary and charge accumulation can occur, leading to a numerical sheath region with spurious electric fields. In order to circumvent this problem, the Poisson equation is solved with an artificial `shielding' density in addition to the prescribed density near the boundary. The shielding density, as shown in \figref{27268_profiles}(b), peaks at the plasma boundary, decaying exponentially over a scale length $\Delta\rn = 0.02$. Sensitivity studies have been carried out to determine the effect of changing the scale length of the shielding density, whereby it was found that increasing this from $\Delta\rn = 0.02$ to 0.03 has little effect on the amplitude profile, which in all simulated cases peaks deeper inside the plasma (at or inside $\rn \sim 0.9$).

As discussed in \secref{synth_data}, generation of the synthetic BES data requires output of the simulated density fluctuations $\dnene(R, Z, t)$ as a function of time over a 2D (radial\footnote{Here the radial coordinate $R$ is defined relative to the symmetry axis of the tokamak.} $R \times$ vertical $Z$) grid encompassing the spatial extent of the measurements. For simulations with adiabatic electrons, the density fluctuations are calculated from the perturbed potential $\varphi$ by assuming a Boltzmann response for the electrons $\dnene \approx e\varphi/T_e$. For the simulations with kinetic electrons, in order to decrease the computational time, only the trapped electrons are treated kinetically and, in this case, $\dnene$ is calculated from the full distribution function assuming a Boltzmann response for the passing electrons. An example of synthetic data from a non-linear simulation is shown in \figref{2d_dnene}.

\section{\label{sec:exp_obs} Experimental measurements}

The ion-scale density fluctuations in MAST plasmas are measured using a 2D imaging BES diagnostic \cite{field_rsi_2012}, which is briefly described in \secref{bes_system}. The characteristics of the turbulence (amplitude, correlation times and lengths and apparent poloidal phase velocity of the density contours) are determined using the correlation analysis described in \secref{corr_anal}, applied to short ($\sim \quant{2}{ms}$) time series corresponding to the equilibrium used for the gyro-kinetic simulations. Care was taken to choose a quiescent time period free of the fast-ion-driven MHD events, with frequencies in the range $\quant{20-200}{kHz}$, which are frequent in NBI heated discharges on MAST. Other data required for our analysis are obtained from the high-resolution profile diagnostics available on MAST: the magnetic pitch angle ($\alpha = \tan^{-1}(\Btheta/\Bphi)$, where $\Btheta$ and $\Bphi$ are the poloidal and toroidal components of the magnetic field) is measured by a 32-channel Motional Stark Effect (MSE) diagnostic \cite{debock_rsi_2008}; ion temperature $T_i$ and toroidal flow velocity $\Uphi = R \omgphi$ are obtained from multi-channel, Charge-Exchange-Recombination Spectroscopy (CXRS) measurements on $\rm{C^{+6}}$ impurity ions with spatial resolution of $\sim \quant{1}{cm}$ \cite{conway_rsi_2006}; and electron temperatures $T_e$ and densities $n_e$ from a 132-channel NdYAG Thomson Scattering system with comparable spatial resolution \cite{scannell_rsi_2010}. The components of the magnetic field vector are obtained from MSE-constrained EFIT equilibrium reconstructions \cite{lao_nf_1985}.

\subsection{\label{sec:bes_system} BES turbulence measurements}

The ion-scale density fluctuations are measured using a 2D imaging BES diagnostic \cite{field_rsi_2012}. This system has a 2D Avalanche Photo-diode Detector (APD) in the form of an 8 radial $\times$ 4 poloidal channel array, which is directly imaged (rather than using optical fibres) onto the heating neutral beam with a resulting nominal spatial resolution of $\sim 2\;\rm{cm}$ radially and poloidally. The Doppler-shifted $\dalpha$ emission from the beam is selected using a band-pass interference filter. The incident light on the APD sensors with a photon flux of $\lesssim 2 \times \quant{10^{11}}{s^{-1}}$ is detected with a SNR $\lesssim 300$ at 2 MHz digitisation rate simultaneously for all channels. If appropriate correlation analysis is used \cite{ghim_prl_2013}, the system is able to detect density fluctuations with relative amplitude $\dnene \gtrsim 0.2\%$ at wave numbers $\abs{k} \leq \quant{1.6}{cm^{-1}}$ and at frequencies up to the Nyquist frequency of 1 MHz.

The actual spatial resolution is degraded below that determined by the imaging properties of the optical system, both due to geometrical effects caused by the finite depth of the line of sight through the beam ($\sim \quant{20}{cm}$), together with field-line curvature and any mismatch between the $\bm B$-field direction and the line of sight, which is optimal when viewing at mid-radius ($R_v \sim \quant{1.2}{m}$). The finite lifetime of the excited deuterium atoms ($\quant{3-10}{ns}$) also causes the $\dalpha$ emission due to instantaneous excitation to be spatially de-localised by $\sim \quant{1-3}{cm}$ in the direction of beam propagation. The spatial response of the BES system is determined using a numerical simulation code \cite{ghim_rsi_2010}, which takes account of these effects in terms of 2D point-spread functions, as discussed in \secref{synth_data}. This spatial response depends on the particular magnetic equilibrium, beam parameters, viewing location and plasma profiles and so has to be calculated explicitly for each measurement.

The $\dalpha$ emissivity of the beam is proportional to the beam density $n_{0}$, the plasma density $n_{e}$ at the observed location and to the relative population of the $n=3$ excited state, which is determined by a collisional-radiative balance between excitation and de-excitation of the beam atoms by collisions with the plasma ions and electrons and by radiative decay. Neglecting any low-frequency (a few kHz) fluctuations in the beam density, the relative density fluctuation level can be determined from the intensity fluctuations using the relation $\dnene = (1/\dlnidlnn)(\delta I/I)$, where $\dne$ ($\dI$) and $n_{e}$ ($I$) are the fluctuating and mean components of the plasma density (intensity of emission), respectively \cite{fonck_rsi_1990}. The differential excitation rate $\dlnidlnn$ is obtained from the data in \refref{hutchinson_ppcf_2002}, where it is calculated using an appropriate collisional-radiative model. This parameter is a weak function of density (decreasing from 0.93 to 0.27 over the density ranging from $n_e = 10^{18}$ to $10^{20} \rm{m^{-3}}$) and a very weak function of temperature.

\begin{figure}[t]
\begin{center}
\includegraphics[width=5.0in]{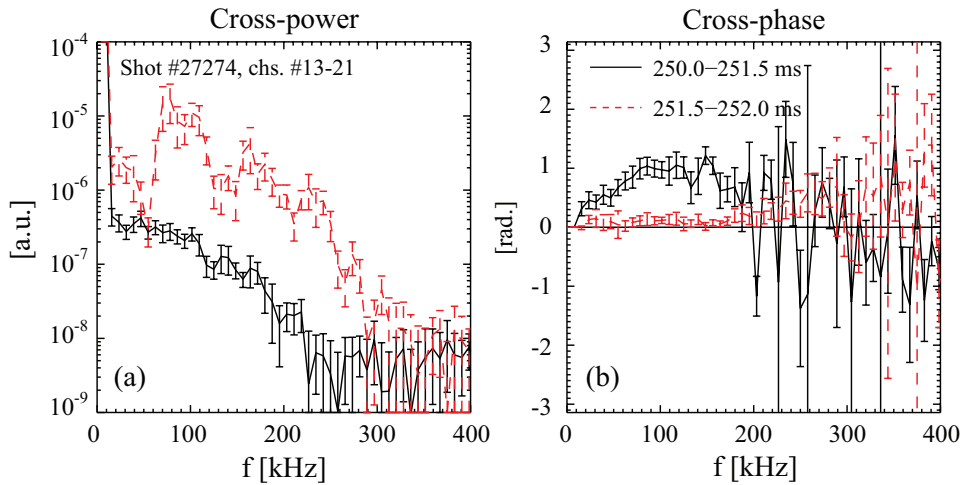}
\caption{\label{fig:power_spectra} Examples of (a) cross-power and (b) cross-phase spectra between two poloidally adjacent channels (\#13 and \#21) of the BES array viewing the location $\rn \simeq 0.82$ during the period used for the correlation analysis ($\quant{250.0-251.5}{ms}$, solid black) and just after, when there is strong MHD activity ($\quant{251.5-252.0}{ms}$, dashed red).}
\end{center}
\end{figure}

The ability of the BES diagnostic to detect density fluctuations due to low-frequency, ion-scale turbulence - the phenomenon of interest in this study - can be compromised by density fluctuations due to fast-ion-induced MHD activity superimposed on the data. Examples of cross-power and cross-phase spectra (between two poloidally adjacent channels at the same nominal radius corresponding to $\rn \simeq 0.82$) of the BES fluctuation data are shown in \figref{power_spectra} for two periods corresponding to the MHD-quiescent phase used for the correlation analysis and a period just after during which there is fast-ion-induced MHD activity. During the MHD-free phase, the turbulent component of the signal decreases to the level of the broad-band noise ($\lesssim 5 \times10^{-8} $) at frequencies $> \quant{200}{kHz}$. The linear increase in the cross-phase with frequency up to $\quant{100}{kHz}$ observed during the first period is expected for broadband turbulence propagating poloidally at constant phase velocity. During the second period, the cross-power increases by an order of magnitude and is dominated by the MHD component, which exhibits a wide range of frequencies due to the sweeping nature of the fast-ion driven `fish-bone' instabilities \cite{chen_prl_1984}. The spatially coherent nature of the MHD-induced fluctuations is manifest in the zero cross-phase over all frequencies below $\quant{200}{kHz}$. In order to avoid the complications imposed by the MHD, we have applied the correlation analysis described in \secref{corr_anal} only to short time periods relatively free of this activity. (In previous work \cite{ghim_prl_2013}, we applied data selection cuts which reject data from periods with strong, coherent MHD activity.)

Another general issue with core fluctuation diagnostics based on BES\color{blue}, which was first raised with reference to the measurements on TFTR \cite{durst_rsi_1992} \color{black} is that strong density fluctuations at the plasma periphery can be imprinted on the beam density, hence superimposing a spatially coherent component onto the observed fluctuations in the plasma core, which is in anti-phase with those at the edge. Methods exist to account for this effect by performing an inversion of an integral transformation, involving the so-called 'beam-transfer function', over the beam path from the edge to the observed location in the plasma, e.g. as described in \refref{zoletnik_ppcf_1998}. Such methods can, however, only be applied if the fluctuation measurements are available simultaneously over the beam path from the edge to the deepest observed location. As the MAST BES system has a fixed detector geometry, it is impossible to observe the edge and core plasma simultaneously, so such techniques cannot be applied to our data. However, the correlation analysis described in \secref{corr_anal} accounts for spatially constant contributions to the measured correlation functions. Let us nevertheless give a simple estimate that shows that the beam-imprinting effect on our measurements is small.

The detection limit for core fluctuations in terms of $\dnene$  imposed by this beam imprinting effect due to a given level of edge fluctuations can be estimated as follows. The $\dalpha$ emissivity from the beam is proportional to the beam density $n_b$, hence fractional changes in the beam density $\dnbnb$ will produce a similar fluctuation in the observed $\dalpha$ intensity and a factor $1/\dlnidlnn$ larger fractional change in the inferred density fluctuation level.  An approximate upper limit to the apparent core fluctuation level $\lp \dnene \rp_{app}$ due to an edge density fluctuation in an edge-localized `shell' of line-integral density $\bar{n}_e = n_e \Delta x$, where $\Delta x$ is the shell's thickness, is $\lp \dnene \rp_{app} \sim \lp \dnene \rp_{edge} \lp S_b/v_b \rp \bar{n}_e / \dlnidlnn$, where $S_b$ is the beam stopping rate coefficient and $v_b$ is the beam-atom velocity (for a $D^0$ beam at $\quant{70}{keV}, v_b/S_b \sim 2 \times \quant{10^{19}}{m^{-2}}$). Estimating the average fluctuation level as $2\%$ over a peripheral shell with $\bar{n}_e \sim \quant{10^{18}}{m^{-2}}$ (equivalent to the region $\rn \geq 0.8$) yields $\lp \dnene \rp_{app} \sim 0.3\%$, which is below the observed fluctuation level in the plasma core inferred from our measurements (see \figref{amp_qi})\footnote{\color{blue}Note that density turbulence has been observed using BES system on the NSTX device in the pedestal region of H-mode plasmas \cite{smith_nf_2013}. Although the amplitudes reported there of $\dnene \sim 1-5\%$ are similar to those we observe in the peripheral region of L-mode plasmas, this turbulence is highly localised to the steep gradient region of the pedestal.\color{black}}. Note that for a Li beam, the penetration depth is an order of magnitude less than for a deuterium beam at comparable energy, hence it is far more important to account for the beam-imprinting effect in the analysis of fluctuation data from BES systems utilising a Li beam \cite{zoletnik_ppcf_1998}.

\subsection{\label{sec:corr_anal} Correlation analysis}

The statistical characteristics of the fluctuations are determined using correlation analysis techniques very similar to those used in \refref{ghim_prl_2013}, to which the reader is referred for further details. The data is first band-pass filtered over the frequency interval $\quant{20-200}{kHz}$ to reject high-frequency noise and the beam fluctuations at a few kHz, which otherwise disturb the correlation analysis \cite{ghim_ppcf_2012}. A matrix of spatio-temporal correlation functions is then calculated according to:
\setlength{\mathindent}{4 pc}
\begin{eqnarray}
\label{eq:corr_def}
\newcounter{eq:corr_def}
\setcounter{eq:corr_def}{1}
\corr\lp x,Z,\Dx,\DZ,\Dt \rp = 
\frac{\lab \dI\lp x, Z, t \rp \dI\lp x+\Dx, Z+\DZ, t+\Dt\rp \rab}{\sqrt{\lab\dI^2\lp x, Z, t\rp\rab \lab\dI^2\lp x+\Dx, Z+\DZ,  t+\Dt\rp\rab}},
\end{eqnarray}
\setlength{\mathindent}{6 pc}
where $x$, $Z$ and $t$ denote the radial and vertical (poloidal) positions and time, respectively, $\Dx$ and $\DZ$ are the radial and poloidal channel separations, $\Dt$ is the time lag and  $\lab \cdot \rab$ denotes a time average over a period of $\sim \quant{2}{ms}$ for the experimental data and over the periods given in \tabref{sim_cases} for the simulated data. Note that we use the notation ($x, y$) for the radial and perpendicular directions relative to the flux surfaces, which for the case of the up/down symmetric, double-null diverted (DND) equilibria considered here are aligned with the ($R, Z$) directions at the horizontal mid-plane.

The auto-covariances $\acov \lp x,Z,\Dt \rp= \lab \dI \lp x,Z,t \rp\,\dI \lp x,Z,t+\Dt \rp \rab$ at $\Dx = \DZ = 0$, contain not only a component due to the true density fluctuations but also photon and electronic noise. The density fluctuation level at each radial location can be obtained from the (noise-subtracted) auto-covariance functions according to: $\dnene \lp x \rp = \lp 1/\dlnidlnn \lp x\rp\rp \cdot \{\sqrt{\acov\lp x, Z, 0 \rp-\acov_{N} \lp x, Z, 0\rp}/I\lp x, Z\rp \}$, where $\acov_{N}$ is the auto-covariance of a signal from a calibration source containing only noise, determined at all 32 locations, then averaged (as denoted by $\{\cdot\}$) over the four poloidally separated channels at the same radial location.

\begin{figure}[t]
\begin{center}
\includegraphics[width=5.0in]{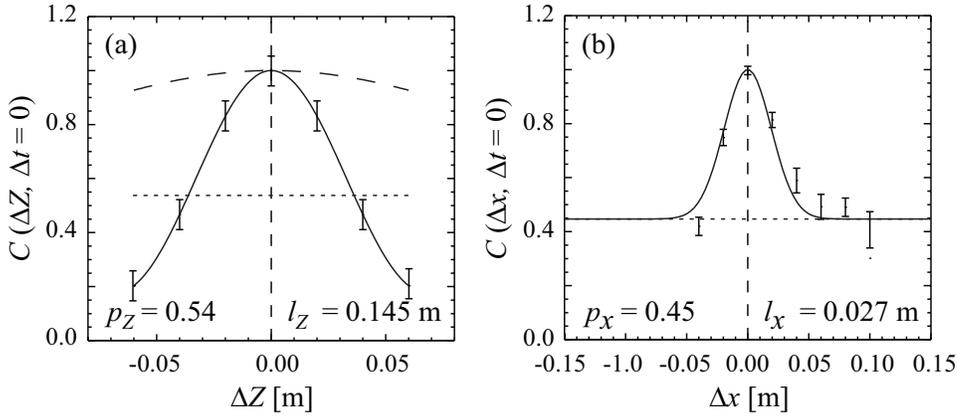}
\caption{\label{fig:corr_functs} (a) Poloidal and (b) radial correlation functions measured at $\rn \simeq 0.82$, showing the measurements (points), fitted functions $f_Z$ and $f_x$ (solid), exponential envelope of $f_Z$ (dashed) and offsets $p_Z$  and $p_x$.}
\end{center}
\end{figure}

The poloidal correlation length $\lZ$ is estimated using data from the four poloidally spaced channels at each radial location by fitting the averaged cross-correlations (over channels with the same $\DZ$ at each radial location $x$), $\bar{\corr} \lp x, \Dx = 0, \DZ, \Dt = 0\rp$ to the function $f_Z\lp\DZ\rp = p_Z + \lp 1 - p_Z \rp \cos \lsb 2\pi \DZ / \lZ \rsb \exp \lsb - \lp \DZ/\lZ \rp^2 \rsb$, where $p_Z$ and $\lZ$ are the two fit parameters. The parameter $p_Z$ effectively subtracts any spatially constant component, usually due to large-scale, global MHD modes. The field-perpendicular correlation length on a given flux surface is then $\ly = \lZ \cos \alpha$, where $\alpha = \tan^{-1}(\Btheta/\Bphi)$ is the pitch angle of the magnetic field. An example of a fitted poloidal correlation function is shown in \figref{corr_functs}(a) for channels at a nominal location of $\rn \simeq 0.82$ for which the polodially constant offset term was $p_Z = 0.54$. The relatively large value of this offset is likely to be due to residual MHD activity, which is never entirely absent during the beam-heated phase.  Note that fitting this offset serves to subtract both any component due to residual, global MHD activity or from edge-induced beam density fluctuations as discussed in \secref{bes_system} above. \color{blue}This technique is essentially equivalent to that discussed in \refref{durst_rsi_1992}, where the contribution due to edge-induced beam density fluctuations was accounted for by subtracting a long-range, spatially constant component from the local cross-correlation functions.\color{black}
 
In contrast to the poloidal correlation functions, the radial correlation functions have a monotonically decaying rather than a wave-like structure. Radial correlation lengths $\lx$ are therefore obtained by fitting the averaged radial cross-correlations (over the four poloidal channels at each radial location $x$), $\bar{\corr} \lp x, \Dx, \DZ = 0, \Dt = 0\rp$ to the function $f_x\lp\Dx\rp = p_x + \lp 1-p_x \rp \exp \lsb - \lp \Dx/\lx \rp^2 \rsb$, where the fit parameters are $\lx$ and the constant $p_x$, again serves to subtract any spatially constant component. An example of a fitted radial correlation function is shown in \figref{corr_functs}(b). The radially constant offset term in this case was $p_x = 0.45$, which is quite close to $P_Z$ obtained above from fitting the poloidal correlation function.

The correlation time $\tc$ is estimated from the temporal cross-correlation functions $\corr\lp\Dx=0,\DZ,\Dt\rp$ between the four poloidally separated channels at each radial location, by finding for a particular $\DZ$ the time delay $\taupeak \lp\DZ\rp$ at which the correlation function has its maximum. The peak values of the correlation functions $\corr_{\rm peak} \lp \taupeak \lp\DZ \rp\rp$ are then fitted with the function $f_{\Dt} \lp \taupeak \lp\DZ \rp\rp = \exp \lsb -\abs {\taupeak \lp\DZ\rp}/\tc \rsb$ to determine $\tc$, where it is implicitly assumed that any influence of the finite parallel correlation length $\lpar$ on $\tc$ can be ignored.

The apparent phase velocity $\Ubes$ of the fluctuations is determined at each radial location using a cross-correlation time delay (CCTD) technique \cite{durst_rsi_1992}, namely it is calculated by making a linear fit to $\DZ \lp \taupeak \rp$. It is found that the observed velocity is dominated by the apparent poloidal motion of field-aligned, elongated ($\lpar/\lxy \gg 1$) eddies through a poloidal plane due to the dominant toroidal rotation of the plasma: as discussed in \refref{ghim_ppcf_2012}, it can be shown that $\Ubes \approx - \Uphi \tan \alpha + \UZ$, where the toroidal velocity $\Uphi \gg \UZ$. This is because any poloidal flows are strongly damped \cite{hinton_pof_1985}, leaving $\UZ$ of the order of the diamagnetic velocity $\Udiai \sim \rhost \vti$.

This type of analysis, characterising the turbulence in terms of a few scalar parameters ($\dnene, \lx, \ly$ and $\tc$) determined from the ensemble-averaged correlation functions, effectively yields these parameters averaged over all density fluctuations at spatial scales larger than the instrumental resolution ($\kper \leq \quant{1.6}{cm^{-1}}$), weighted by the RMS value of the density fluctuations $\lab\lp\dnene\rp^2\rab^{1/2}$. These parameters are therefore representative of the fluctuations at the dominant `outer', energy-containing scale of the turbulence \cite{barnes_prl_2011_107}. In order to calculate meaningful ensemble averages, the time averaging must be done over many correlation times and wave periods, i.e., over a sample of duration $\Dt_{avg} \gg 2\pi/\omgsti$, where $\omgsti \sim \vti\rhoi/\lp\ly\LTi\rp$ is the ion-diamagnetic frequency and $\LTi \equiv (\nabla T_i/T_i)^{-1}$ is the scale length of the ion temperature gradient. Taking the fiducial value $\omgsti/2\pi \sim \quant{10}{kHz}$, sample periods of order 1\;ms are required.

\subsection{\label{sec:synth_data} Synthetic BES data}

For the comparisons between simulations and measurements to be meaningful, `synthetic' data, analogous to the real BES data, is generated from the simulated density fluctuations and analysed using the same correlation techniques as those used for the experimental measurements. The synthetic data is generated using a synthetic `diagnostic', which mimics the physical characteristics of the measurement technique. The effect of applying this processing to the simulated data is firstly to spatially average over fluctuations at smaller scales than that of the instrumental resolution and secondly to impose a lower limit to the fluctuation amplitude below which the signal is dominated by noise.

In order to generate the synthetic data, the characteristics of the BES system have to be quantified in terms of its spatial resolution, sensitivity and noise properties. These are determined using a numerical simulation, described in more detail in \refref{ghim_rsi_2010}, which accounts for the beam absorption and emission, magnetic field and viewing geometry. Assuming that the turbulent eddies are highly elongated along the field, $\lpar /￼ \ly \gg 1$, the spatial response of each channel $(i, j)$ can be quantified in terms of 2D point-spread functions (PSFs), $\wp_{ij} (R, Z)$, which are effectively cross-sections for the excitation of the $\dalpha$ beam emission per unit area of the field of view at the focal plane at the beam.

The PSFs depend on the beam properties, magnetic equilibrium and plasma profiles, as well as on the diagnostic parameters (e.g., on the APD bias voltage and radial viewing location) and so have to be calculated explicitly for each simulated equilibrium and diagnostic setting. The validity of this procedure of utilising 2D PSFs to represent the spatial response relies on the assumption that the turbulence is to a good approximation two-dimensional, specifically that its parallel correlation length $\lpar \gg \dlbeam$, where $\dlbeam \sim \quant{20}{cm}$ is the line-of-sight path length through the beam. If the parallel correlation length $\lpar \sim \Lambda$, where $\Lambda \sim \pi r (B/\Btheta) \gg \quant{1}{m}$ is the connection length at the low-field side of the plasma \cite{ghim_prl_2013, barnes_prl_2011_107}, then this approximation is well satisfied.

\begin{figure}[t]
\begin{center}
\includegraphics[width=6.0in]{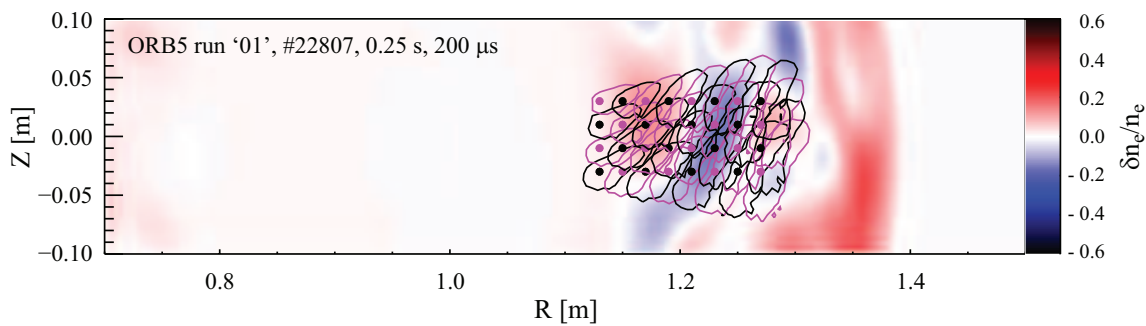}
\caption{\label{fig:2d_dnene} Relative density fluctuations ($\dnene$) from a non-linear simulation of a MAST equilibrium (\#22807 at 0.25\,s) with the nominal viewing locations ($\bullet$) and $1/e$ contours of the respective PSFs of the BES diagnostic superposed.}
\end{center}
\end{figure}

Generation of synthetic BES data from the non-linear simulations requires a 2D map of the relative density fluctuations $\dnene$ over a poloidal cross section (i.e. at fixed toroidal angle) as a function of time on a regular ($R, Z$) grid covering the region of the plasma viewed by the BES diagnostic. With three viewing locations used to measure a full radial profile, the BES measurements cover a radial range of $0.95 \leq R \leq \quant{1.45}{m}$, while the vertical extent of the measurements is $-0.1 \leq Z \leq +\quant{0.1}{m}$. Relative density fluctuations from a NEMORB simulation over such a radial stripe are shown in \figref{2d_dnene} with the PSFs of the BES system superimposed. Because the spatial extent of each channel is of order of a few cm, a resolution of $\DR \sim \DZ \sim 0.5\;\rm{cm}$ is sufficient for the simulated data. The sampling rate of the BES diagnostic is 2 MHz, hence the sample period for the simulated data is also chosen to be $\Dt \sim \quant{0.5}{\mu s}$.

As well as the relative density fluctuation data, some other data is required for the synthetic data generation. Calculation of the rate of $\dalpha$ photons incident on each sensor channel $\Gamma_{ij}(t)$ requires the mean density $n_{e0}(R, Z)$ (which does not need to be a function of time as the duration of the simulation is much shorter than the timescale of the profile evolution) and the simulated relative density fluctuation $\dnene (R, Z, t)$ data. The photon rate $\Gamma_{ij}(t)$ can then be calculated from the following expression:
\begin{eqnarray*}
\label{eq:da_flux}
\Gamma_{ij} = \int \int \wp_{ij} \lp R, Z \rp n_{e0} \lp R, Z \rp \lsb 1 + \dlnidlnn \lp \delta n_e \lp R, Z, t \rp / n_e \rp \rsb dR\,dZ,
\end{eqnarray*}
where the differential excitation rate $\dlnidlnn = \lp\delta I/I\rp/\lp\dnene\rp$ was discussed in \secref{bes_system} and the integral is performed over the $(R, Z)$ region covered by the simulated data. 

The statistical photon noise within the digitisation period is simulated as follows. The number of detected photons in a sample period $\Dt = 1/\fBW$, where $\fBW \sim 5 \times 10^{5}\; \rm{Hz}$ is the bandwidth of the pre-amplifiers, is $N_{ij} = \Qeff \FT\,\Gamma_{ij}/\fBW$, where $\Qeff \sim 0.85$ is the quantum efficiency of the detectors and $\FT \sim 0.25$ is a factor to take account of the transmission of the optics. Pseudo-photon noise is then added according to $N_{ij}^\star(t) = N_{ij} + \Re_N \cdot \sqrt{\FN N_{ij}(t)}$, where $\Re_N$ is a random number from a normal distribution of unit standard deviation. The coefficient $\FN$ is the excess noise factor due to the amplification process in the APD, which is given by $\FN = \GAPD^{\kappa} \sim 2$, where $\kappa \sim 0.3$ is the excess noise exponent due to the amplification processes in the APD, and $\GAPD$ is the gain of the APD ($\GAPD \sim 10$ at the bias voltage used for the measurements of 310\;V). The output voltage is then calculated from $V_{ij}^{\star}(t) = \Imamp\,\GAPD\,e\,N_{ ij}^{\star}\,\fBW + \Re_N\,\sigma_V$, where $e$ is the electron charge and $\Imamp = 3.4 \times 10^6\;\rm{V/A}$ is the trans-impedance of the amplifiers. The second term is the electronic noise of the amplifier, also simulated by adding a normally distributed random noise voltage with standard deviation $\sigma_{V} = \quant{2.5}{mV}$. Typically, the output voltage is $\lesssim \quant{1.2}{V}$, hence the SNR due to the amplifier noise alone is $\lesssim 480$ and the overall SNR including the photon noise is $\lesssim 250$.

The procedure followed to generate the turbulence characteristics for the simulated turbulence data can be summarised as follows: PSFs for each of the three BES view radii are generated for the specific equilibrium and beam parameters used in the experiment; $\dnene(R, Z, t)$ data is written out from the simulation over the required domain; the synthetic BES data are generated for each available PSF; finally, the synthetic data is analysed using the same correlation analysis as for the real BES data, which was described in \secref{corr_anal}.

\section{\label{sec:nl_comp} Comparison of non-linear simulations with measurements}

\begin{figure}[t]
\begin{center}
\includegraphics[width=5.0in]{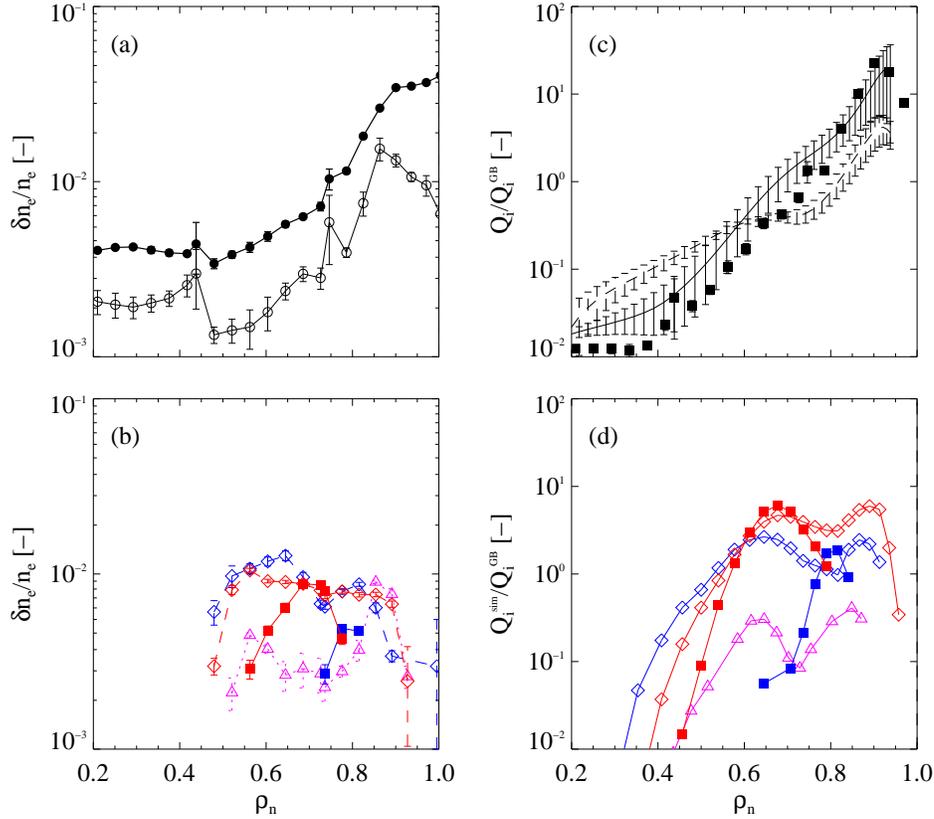}
\caption{\label{fig:amp_qi} Profiles of the normalised fluctuation amplitude $\dnene$ from (a) experiment (\#27268) during the beam phase at 0.25 s ($\bullet$) and after beam cut-off at 0.3 s ($\circ$); and (b) from the five gyro-kinetic simulation cases in \tabref{sim_cases}, where the plot symbols are also defined. Profiles of the gyro-Bohm normalise ion heat fluxes $\Qiexp/\QiGB$: (c) from experiment as determined by TRANSP calculations (solid line), together with the  normalised neo-classical value $\QiNC/\QiGB$ (dashed line) and as estimated from the measured turbulence amplitude $\Qiturb/\QiGB$ ($\blacksquare$); and (d) $\Qisim/\QiGB$ from the gyro-kinetic simulations.}
\end{center}
\end{figure}

Results of applying the above procedure to the five non-linear NEMORB simulations of the MAST L-mode discharge \#27268 discussed in \secref{nonl_gk_sim_27268} are presented below and the simulated turbulence characteristics are then compared with experimental measurements. Profiles of the experimental and predicted fluctuation amplitudes $\dnene$ and ion heat fluxes $Q_i$ are shown in \figref{amp_qi} as a function of normalised radius $\rn \equiv \psinh$. Because the radial extent of the BES array covers only one third of the outboard plasma radius of $a \sim \quant{0.5}{m}$, the experimental data in this figure are taken from three similar discharges, with nominally the same equilibrium and kinetic profiles, but with the BES viewing position set to three different radial locations (\#27272, 1.05\;m, \#27268, 1.2\;m and \#27274, 1.35\;m). The correlation analysis is applied to a $\sim \quant{2}{ms}$ data sample at the time $\sim \quant{0.25}{s}$, corresponding to the equilibrium used for the simulations (note that this is $\quant{50}{ms}$ before the time of the beam cut-off at 0.3\;s).

\subsection{\label{sec:comp_amp_qi}Turbulence amplitudes and ion heat fluxes}

As shown in \figref{amp_qi}\,(a), the experimental fluctuation level in the core is $\sim 0.4\%$, which is significantly above that due to background light emission of $\sim 0.2\%$ and increases strongly to reach $\sim 4\%$ towards the periphery. The background fluctuation level is estimated by normalising the fluctuation amplitude just after the beam cutoff $\delta I_{bg}$ to the signal level $I_{B}$ just before the beam cut. Hence, $\delta I_{bg}/I_{B}$ should be representative of the background fluctuation level during the beam phase provided this does not change significantly between the time of observation and the beam cut-off. Spectral observations show that the background emission is predominantly so-called FIDA (fast-ion $\dalpha$ emission) from re-neutralised beam ions, which increases towards the plasma periphery where the neutral density is highest. Note that, for $\rn \gtrsim 0.5$, the observed fluctuation level in the core is larger than the detection limit estimated in \secref{bes_system} as $0.3\%$ due to edge fluctuation induced beam density fluctuations.

The experimental heat flux, determined from transport analysis \cite{field_nf_2011} performed using the TRANSP code \cite{hawryluk_cec_1980}, is shown in \figref{amp_qi}\,(c). The uncertainties in the absolute levels of $\Qiexp$ are quite large due to the difficulty in quantifying the level of anomalous fast-ion redistribution and losses arising from fast-ion driven MHD activity, which is particularly strong during the phase with both on-axis directed NBI heating beams for which this comparison is made. Outside $\rn \sim 0.2$, the heat flux is approximately constant at $\Qiexp \sim \quant{0.02}{MWm^{-2}}$ and the value normalised to the gyro-Bohm level $\Qiexp/\QiGB$, where $\QiGB = n_i T_i\,\vti\,\lp\rhoi/R\rp^2$, increases by over three orders of magnitude from $\sim10^{-2}$ in the core to $\sim 10$ in the plasma periphery. In the outer region of the plasma, $\Qiexp$ is up to a factor $\sim 5$ larger than the neo-classical heat flux $\QiNC$ determined from NCLASS \cite{houlberg_pop_1997}, but it is below the neo-classical level in the core, where the power balance is particularly sensitive to assumptions made about the anomalous fast-ion diffusion.

Determining the turbulent heat flux in principle requires knowledge of the density, temperature and potential fluctuations and cross-phases between them \cite{white_pop_2010}. Measurements of potential fluctuations in the core plasma are challenging and rarely available \cite{ido_rsi_2008} and those of ion-temperature fluctuations even more so \cite{evensen_nf_1998}. In the absence of such measurements on MAST, the gyro-Bohm-normalised turbulent ion heat flux can be estimated from the simplified relation:  $\Qiturb/Q_i^{GB} \sim \ky \rhoi \lp T_e/T_i\rp^2 \lp\dnene\rp^2/\lp\rhoi/R\rp^2$, where the Boltzmann relation between $\varphi$ and $\dnene$ is assumed and any possible reductions due to the cross-phase are ignored. As can be seen in \figref{amp_qi}\,(c), this simplified estimate agrees well with that obtained from power balance at the periphery $\rn \gtrsim 0.8$,  but underestimates $\Qiexp$ by a factor $\sim 2$ over the rest of the profile. Interestingly, measurements on TFTR \cite{evensen_nf_1998} found that in plasmas with dominant ITG turbulence, $(\delta T_i/T_i)/(\dnene) \sim 2$ over the outer half of the plasma radius, so an appreciable fraction of the ion heat flux might arise from enhanced ion temperature fluctuations in regions with dominant ITG turbulence.

\begin{figure}[b]
\begin{center}
\includegraphics[width=2.5in]{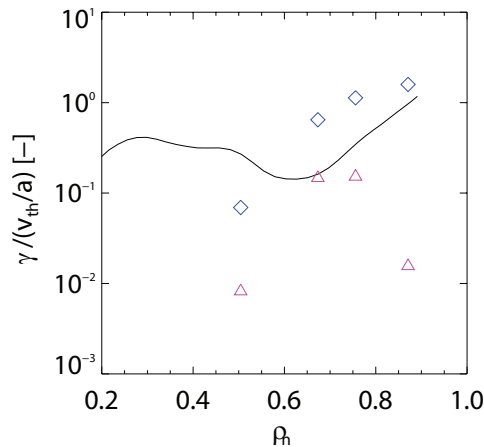}
\caption{\label{fig:gegm}Profiles of the normalised $\ExB$ shearing rate $\gEbar$ (solid line) calculated by TRANSP for the equilibrium from \#27268 at $\quant{0.25}{s}$ and the normalised linear growth rates $\gm/(\vti/a)$ calculated by GS2 for cases I (\textcolor{magenta}{$\triangle$} -- with adiabatic electrons, no flow) and II (\textcolor{blue}{$\diamond$} -- kinetic electrons, no flow or collisions) for the similar equilibrium from \#22807 at $\quant{0.25}{s}$. \color{black}}
\end{center}
\end{figure}

The normalised $\ExB$ shearing rate, $\gEbar = \gE/(\vti/a)$, is shown in \figref{gegm}, together with maximum growth rates $\gm$ from linear GS2 calculations (in this case from the similar equilibrium of discharge \#22807 at 0.25\;s) for two cases (\rm{I} and \rm{II}) without sheared flow or collisions. While for case \rm{I} (adiabatic electrons), $\gE > \gm$ over the whole profile, for case \rm{II} (kinetic electrons), $\gE > \gm$ only in the region with strong flow shear inside $\rn \sim 0.5$, which is consistent with the turbulence being suppressed by the flow shear there.

As shown in \figref{amp_qi}\,(d), the simulation with adiabatic electrons but without sheared flow (\rm{I}) results in an ion heat flux $\Qisim$ up to an order of magnitude below $\QiNC$ and a correspondingly small fluctuation amplitude. Kinetic treatment of the trapped electrons only, again for a case without flow (\rm{II}), results in values of $\Qisim$ up to an order of magnitude larger than $\Qiexp$ at mid-radius, while in the core and peripheral regions $Q_i$ is underpredicted. The fluctuation level $\dnene$ exhibits similar behaviour. The underprediction in the core is consistent with the previous linear studies \cite{saarelma_ppcf_2012}, which showed the region where $\hat{s} \leq 0$ to be linearly stable. The effect of including equilibrium flow shear (\rm{III}) in the simulations with kinetic electrons is not dramatic, with a slight decrease in $Q_i^{sim}$ compared to case \rm{II} in the core where the flow shear is strongest, but an increase in the peripheral region, where $\Qisim$ approaches the experimental level -- the corresponding changes in $\dnene$ are rather slight. 

Introducing collisions (\rm{IV} and \rm{V}) reduces the drive from trapped electrons, hence decreasing the level of turbulence and the heat flux. The effect of this is to reduce the radial extent over which there is significant turbulence, in the case \rm{IV} (without flow) to a limited region around $\rn \sim 0.8$, while introducing sheared flow (\rm{V}) broadens and shifts the unstable region inwards (compared to case \rm{IV}) to $\rn \sim 0.7$. In case \rm{V}, with the most complete physics, the peak value of $\Qisim$ exceeds $\Qiexp$ somewhat in the mid-radius region, while the fluctuation level is close to that measured. In the peripheral region, both of these cases underpredict $\dnene$ and $\Qiexp$ (this is not surprising because in NEMORB the fluctuation level is forced to be zero at the plasma boundary).

It is perhaps surprising that introducing sheared flow into case \rm{V} actually increases the predicted heat flux by almost an order of magnitude at mid-radius, when sheared flow is normally expected to suppress the turbulence. Currently, we do not have an explanation for this result, which would necessiate at least performing further global, linear calculations for the specific equilibrium. \footnote{A possible explanation is suggested by the the results of the linear stability calculations presented in Fig. 4 of \refref{saarelma_ppcf_2012}, where it was found that introducing moderate levels of co-current flow shear $\gEbar \sim \rm{\order(0.1)}$ into the global simulations modestly increased the maximum growth rates $\gm$ relative to the case without shear flow. It can be seen from \figref{gegm} that $\gEbar \lesssim 0.2$ at mid radius, so such an increase in growth rates there might explain the increased heat flux compared to case \rm{IV}. The explanation of this effect suggested in \refref{saarelma_ppcf_2012} is that the sign of the toroidal flow determines whether the diamagnetic contribution to the net perpendicular flow shear $\gE$ either enhances or reduces that due to the sheared equilibrium flow, hence either stabilising or destabilising the turbulence \cite{kishimoto_ppcf_1999}.}

The fact that, particularly in the cases with collisions, the simulated turbulence has significant amplitude in relatively restricted radial regions indicates that the gradient is marginally close to the non-linear critical gradient $(\RLTi)_{crit}$ to excite turbulence. This was borne out by the non-linear studies with NEMORB \cite{saarelma_ppcf_2012} carried out for the similar equilibrium of discharge \#22807 (see \secref{nonl_gk_sim}), which showed that relatively small increases in $\RLTi$ above the experimental value could produce large increases in the turbulent heat flux. Although such calculations have not been carried out for the equilibrium used for these studies, comparison of similar non-linear simulations for both equilibria show $\RLTi$ to be even closer to marginality in the equilibrium from \#27268 due to the relatively smaller predicted heat fluxes.

Using our database of equilibrium and turbulence data, primarily from MAST L-mode discharges, it is found \cite{field_iaea_2012, ghim_nfl_2013} that $\RLTi$ exhibits a dependence on flow shear and magnetic geometry, specifically on the parameters $q/\epsilon$ and $\gEbar$. This dependence shows remarkable similarity to a numerically predicted `manifold' of $\RLTi(q/\epsilon, \Uphipr)$ (where $\Uphipr = dR\omega/dr/(\vti/R) \approx \lp q/\epsilon \rp \gEbar$), required for the excitation of marginally unstable ITG- or PVG-driven turbulence \cite{highcock_prl_2012}. This observation is further evidence that the $\RLTi$ is generally close to marginal stability in MAST L-mode plasmas. A consequence of this closeness of $\RLTi$ to marginality is that the use of the heating operator in NEMORB to match the predicted $T_i$ profile to that prescribed will yield heat fluxes which are very sensitive to experimental uncertainties. Hence, it is not surprising that there is not better agreement between predicted profile of $\Qisim$ and experiment.

The shortfall in the predicted level of turbulence and heat flux $\Qisim$ in the peripheral region is a clear deficiency of these non-linear NEMORB simulations, which probably arises from the necessary boundary conditions (see \secref{numerics}) that suppress any turbulence at the plasma edge. As mentioned in the introduction, such a shortfall has also been found in local, non-linear simulations of DIII-D L-mode discharges using the global code GYRO \cite{holland_jop_2008, holland_pop_2009, rhodes_nf_2011}. In these studies, although good agreement was found in the plasma core between simulated and experimental ion and electron heat fluxes and turbulence characteristics, a systematic shortfall in the heat flux and fluctuation levels was observed at $r/a \sim 0.8$ \cite{holland_jop_2008}. More recently, concerted efforts have been made by several groups to either understand or resolve this discrepancy \cite{howard_pop_2013, jenko_itpa_2013} and currently this apparent shortfall appears not to be ubiquitous.

\subsection{\label{sec:comp_lcxy}Perpendicular and radial correlation lengths}

\begin{figure}[b]
\begin{center}
\includegraphics[width=5.0in]{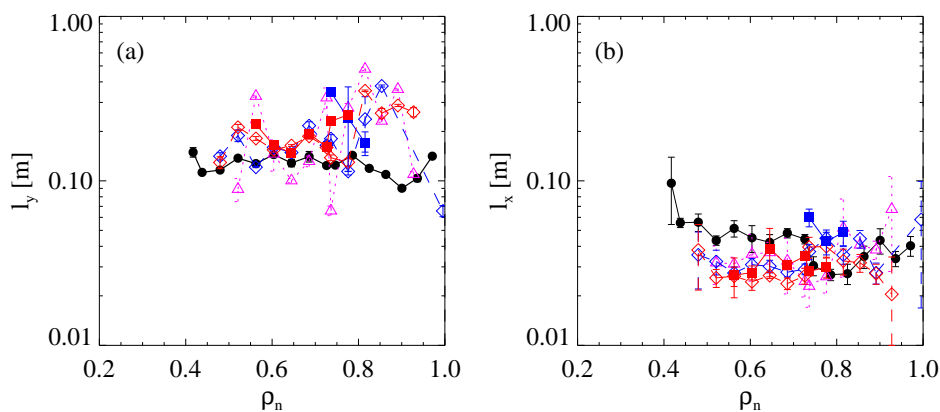}
\caption{\label{fig:lcxy} Profiles of (a) perpendicular $\ly$ and (b) radial $\lx$ correlation lengths determined from experiment ($\bullet$) and from the simulations for the five cases, where the symbols are defined in \tabref{sim_cases}.}
\end{center}
\end{figure}

As shown in \figref{lcxy}, the measured perpendicular correlation length $\ly \sim \quant{10-20}{cm}$ is approximately constant, whereas the radial correlation length $\lx \sim \quant{2-6}{cm}$ decreases somewhat with radius. The observed mean anisotropy $\ly/\lx \sim 3 \pm 1.4$ is consistent with the findings of \refref{ghim_prl_2013}, where it was calculated over a much larger database and was $\sim 5 \pm 2$. The corresponding values of $k_{x,y}\rho_i$ (where $k_{x,y} = 2\pi/\lxy$) are $\ky\rhoi \sim 0.3-1.0$ and $\kx \rhoi \sim 1-2$, generally decreasing with increasing radius. \footnote{Similar values of $\ly$ were already reported in the study of inter-ELM, pedestal turbulence in NSTX \cite{smith_pop_2013} but in a very different turbulence regime to the L-mode plasmas studied here.}

In all of the simulations, the perpendicular correlation length $\ly$ is comparable to that observed over most of the profile except in the periphery ($\rn > 0.7$), where the simulated turbulence may be impacted by boundary conditions. In the simulations, the radial correlation lengths $\lx$ agree within a factor $\sim 2$ with the observed values, although the degree of anisotropy in the simulations is somewhat larger. Note in this context that a related NEMORB study \cite{hill_ppcf_2012} of the dependence of the linear stability of ITG modes on sheared flows showed the anisptropy $\ly/\lx$ of the unstable modes generally increases with the shearing rate.

\subsection{\label{sec:comp_tauc} Correlation times}

\begin{figure}[t]
\begin{center}
\includegraphics[width=2.5in]{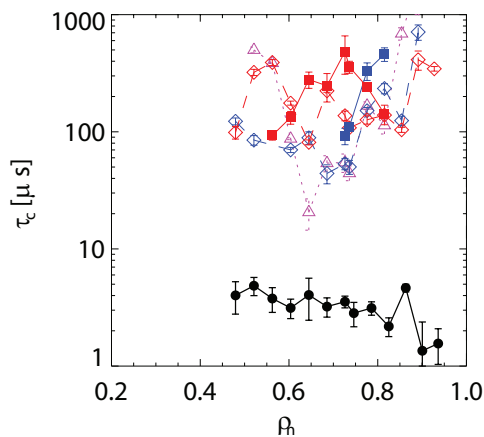}
\caption{\label{fig:tauc} Profiles of the correlation time $\tc$ determined from experiment ($\bullet$) and from the NEMORB simulations for the five cases, where the symbols are defined in \tabref{sim_cases}.}
\end{center}
\end{figure}

The measured correlation times, shown in \figref{tauc}, are rather short and in the range $\tc \sim \quant{1-6}{\mu s}$. One of the most striking of our observations is that in all of the simulations, the correlation times $\tc$ are almost two orders of magnitude longer than those measured. Note that the values of $\tc$ for the simulation cases in \figref{tauc} had to be calculated using synthetic data generated \emph{without} adding noise. This is because, when the turbulence amplitude is weak, the determination of $\tc$ is particularly subject to the effect of noise on the correlation functions and not many valid data points remained for cases \rm{IV} and \rm{V}. Longer-duration simulations would help overcome this problem but prohibitively long computational times would be required. Note, however, that the derived turbulence characteristics were largely insensitive to the addition of noise.

It was shown in \refref{ghim_prl_2013} that the measured $\tc$ are comparable with various linear time scales. These time scales are: the drift-wave time $\tau_{*, i}^{-1} \sim \omgsti$, the parallel streaming time $\tst^{-1} \sim \vti/\Lambda$ (thermal ion transit time over the parallel connection length $\Lambda$) and the magnetic drift time $\tM^{-1} = \vti\rhoi/\lp\lx R\rp$ (perpendicular magnetic drift time over a radial correlation length $\lx$). This observation is consistent with the turbulence being `critically balanced' \cite{barnes_prl_2011_107} and implies that the turbulence is anisotropic in the poloidal plane with $\ly/\lx \sim R/L_*$, where $L_*$ is taken as the shorter of $\LTi$ and $\Lne$ \cite{ghim_prl_2013}. It was also found that the observed correlation times $\tc$ are always shorter than or comparable to the $\ExB$ shearing time $\tsh \sim 1/\gE$, i.e., $\tc \leq \tsh$, which implies that equilibrium $\ExB$ shear does not always determine $\tc$ in the experiment.

\begin{figure}[t]
\begin{center}
\includegraphics[width=6.0in]{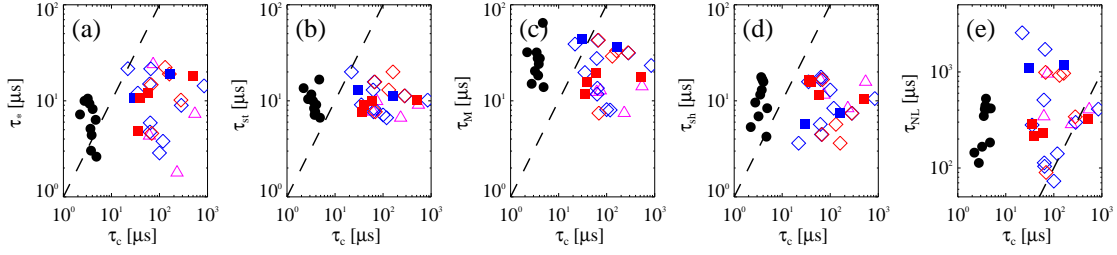}
\caption{\label{fig:tau_comp} Comparison of various time scales with the correlation time of the turbulence $\tc$ from experiment ($\bullet$) and the five simulation cases, for which the symbols are defined in \tabref{sim_cases}: (a) the linear drift-wave time $\tau_*$, (b) the parallel streaming time $\tst$,  (c) the magnetic drift time $\tM$, (d) the shearing time $\tsh$ and (e) the non-linear time $\tnlnz$ estimated from the turbulence characteristics. The dashed lines show $y = x$.}
\end{center}
\end{figure}

In \figref{tau_comp}, the correlation times $\tc$ for the L-mode experimental data considered here, together with those from the five NEMORB simulations, are compared with these time scales. For this experimental data, although $\tau_{*,i} \sim \tst \sim \tM \sim \tsh$, the observed $\tc$ are usually somewhat shorter: the mean logarithms of ratios of these timescales for the experimental data are: $\lab \logten \lp \tstar/\tc \rp \rab = 0.54 \pm 0.61$, $\lab \logten \lp  \tst/\tc \rp \rab = 0.47 \pm 0.37$,  $\lab \logten \lp \tM/\tc \rp \rab = 0.51 \pm 0.52$ and $\lab \logten \lp \tsh/\tc \rp \rab = 0.31 \pm 0.50$. This is consistent with strong, critically balanced turbulence. In the case of the simulation results, the derived $\tc$ are substantially longer than any of the linear time scales. This indicates that in these simulations, the turbulence is weak (whereby $\tc\,\omgsti >> 1$), in contrast to the turbulence measured in the experiment, which appears to be strong ($\tc\,\omgsti \sim 1$). \footnote{Note, however, that the large spread of values of these timescales manifest in \figref{tau_comp} means that, while they clearly are all of the same order, this limited dataset does not provide strong evidence that they are balanced in the same way as was found in \refref{ghim_prl_2013} - there a much larger dataset was used and it was possible to infer that   $\tc \sim \tstar \sim \tst \sim \tM$ held even though local equilibrium parameters changed over a relatively broad range. What we see from the present dataset is in fact just sufficient to confirm that all the linear timescales remain roughly comparable to the general gyrokinetic sound time $a/\vti$. The key finding is that the measured nonlinear timescales associated with the density fluctuations are much longer than $a/\vti$ - and so are the $\tc$ found in the simulations.}

The de-correlation of the turbulence by the fluctuating potential from the drift waves $\phidw$ is characterised by the non-linear time $\tnl^{-1} \sim \vti\rhoi\phidw/\lp\lx\ly\rp$, which can be estimated from the turbulence characteristics by assuming $\dnene$ and $\phidw$ are related through the Boltzmann response. Under this assumption, $\phidw$ does not include any contribution from the trapped-electron response or toroidally and poloidally symmetric zonal flows, so this timescale is denoted $\tnlnz$ and given by $\lp\tnlnz\rp^{-1} \sim \vti\rhoi/\lp\lx\ly\rp \cdot \lp T_e/T_i\rp \dnene$. In \refref{ghim_prl_2013}, it was found that the measured correlation times $\tc$ were generally much shorter than $\tnlnz$ and it was conjectured that this could be consistent with a strong zonal component to the turbulence $\phizf$ contributing dominantly to its de-correlation (the ratio $\tc/\tnlnz$ was also found to increase with collisionality, suggesting that the ratio of the zonal to the drift-wave component of the turbulent amplitude $\phizf/\phidw$ increases with decreasing collisionality).

In the experiment considered here, we also find that the non-linear time $\tnlnz$ is always substantially longer than $\tc$ (see \figref{tau_comp}(e)), the mean logarithmic ratio being $\lab \logten \lp \tnlnz/\tc \rp \rab \sim 2.0 \pm 0.5$. In contrast, in our simulations, the non-linear time $\tnlnz$ is mostly comparable to $\tc$ (e.g. for case \rm{V}, $\lab \logten \lp \tnlnz/\tc \rp \rab \sim 0.2 \pm 1.0$), which indicates that the simulated turbulence is de-correlated by the drift-wave potential fluctuations $\phidw$ and a dominant component from zonal flows need not be invoked to explain the correlation times. Note that with the damping rate for ion-scale zonal flows scaling as $\sim \lp \kpri \rp^2 \nuii$ (perpendicular ion viscosity), non-linear simulations of long duration ($> \quant{1}{ms}$) would be required to capture correctly the zonal-flow dynamics.

\subsection{\label{comp_vpol} Poloidal propagation velocities}

The interpretation of the poloidal propagation velocity of the density patterns observed with the BES system $\Ubes$ is discussed in some detail in \refref{ghim_ppcf_2012}. This motion is not that of the poloidal plasma flow $\UZ$ but arises largely from the dominant toroidal flow $\Uphi >> \UZ \sim \mathcal{O}(\rhost)$ of the field-aligned, elongated `eddies' moving through the radial-poloidal focal plane. In this case, it is evident from simple geometry that this apparent velocity $\Uapp \approx - \Uphi \tan \alpha$, where $\alpha$ is the pitch angle of the local magnetic field line. This expression can also be derived from the continuity equation as in \refref{ghim_ppcf_2012}, where various terms $\sim \mathcal{O}\lp\rhost\rp$ were neglected, including the net, temperature-gradient-driven poloidal flow of the bulk ions, $\UZ \sim \vti \rhoi/\LTi$ \cite{kim_pfb_1991}. Taking $\Uphi$ to be the velocity of the $C^{6+}$ ions measured by the CXRS system $\Uphi = \UphiC$, we can also expect differences between this velocity and that of the $D^+$ ions $\UphiD$ of $\mathcal{O}\lp\rhost\rp$ \cite{kim_pfb_1991}.

\begin{figure}[b]
\begin{center}
\includegraphics[width=5.0in]{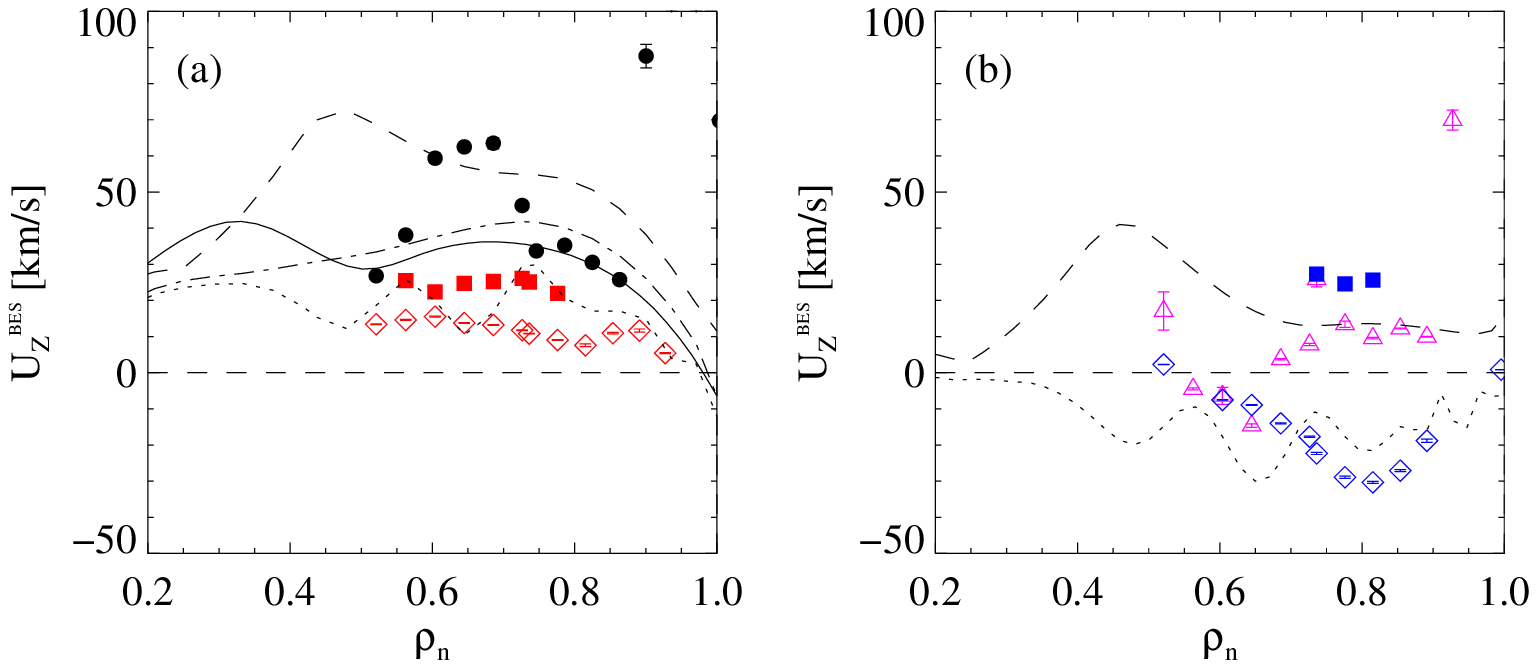}
\caption{\label{fig:vpol} Profiles of the apparent poloidal velocity $\Ubes$ of the fluctuating density patterns for (a) cases with equilibrium flow, including from experiment ($\bullet$) and (b) for cases without equilibrium flow, for which it is assumed that $E_r = 0$, i.e. $\UZExB = 0$. The symbols for simulated cases are defined in \tabref{sim_cases}. Also shown are: $\UappC = -\UphiC \tan{\alpha}$ (solid), the poloidal projection of the $\ExB$ velocity $\UZExB$ (dot-dashed) and the electron- and ion diamagnetic velocities with respect to $\UZExB$, i.e., $\UZExB + \UZ^{\rm dia, \it i,e}$ (dotted and dashed lines respectively).}
\end{center}
\end{figure}

The observed velocity $\Ubes$ can also be considered as the poloidal projection of the field-aligned density patterns propagating with velocity $\Uperp$ perpendicular to $\bm{B}$, i.e. $\Uapp \sim \Uperp / \cos{\alpha}$, the parallel velocity component not contributing to any apparent poloidal motion. Neglecting the diamagnetically small net poloidal plasma flow, this can be related to $\Uphi$ as $\Uapp \sim - \Uphi \tan{\alpha}$, which is the same relation as above. The perpendicular $\ExB$ velocity $\UExB = E_r/B$, where $E_r$ is the radial electric field component, vanishes in the moving frame of the plasma, so any residual perpendicular flow relative to $\UExB$ represents the phase velocity of the density patterns in the plasma frame due to all the $\mathcal{O}(\rhost)$ effects, as discussed in \refref{ghim_ppcf_2012}.

The observed propagation velocity $\Ubes$ is shown in \figref{vpol}(a) together with that calculated from the expression $\UappC = - \UphiC \tan \alpha$ using the $C^{6+}$ velocity from CXRS, where it can be seen that the fluctuations propagate mostly in the positive, ion-diamagnetic direction relative to $\UappC$. The poloidal projection of the $\ExB$ velocity, $\UZExB = -\UExB/ \cos{\alpha}$ is also shown in \figref{vpol}(a). This is determined using the $E_r$ profile from TRANSP, which is calculated using the radial force balance $E_r =  dP_i/dr / \lp e n_i Z_i \rp - \UphiD \Btheta + \UZ^{\lp i\rp} \Bphi$. The components of the $D^{+}$ velocity are calculated in TRANSP from the measured $C^{6+}$ velocity using neo-classical expressions in the NCLASS package \cite{houlberg_pop_1997}.

It can be seen from \figref{vpol}\,(a) that $\UZExB$  is close to $\UappC$. This can be explained by the approximate cancellation of the diamagnetic contribution to $\UExB$ and the difference velocity between the  $D^+$ and the $C^{6+}$ ions, which can be approximated as $\UphiD - \UphiC \sim -U^{\rm dia, \it i} /\sin{\alpha}$. Considering both of these effects it can be shown that $\UZExB \approx \UappC$. In \figref{vpol}, we also plot the sum $\UZExB + \UZ^{\rm dia,\it i,e}$, where the poloidal projections of the ion and electron diamagnetic velocities are $U_Z^{\rm dia,\it e,i} \sim T_{e,i}/\lp L_{P_{e,i}} B\rp/\cos\alpha$ and the pressure-gradient scale lengths $L_{P_{e,i}}$ are estimated using the TS and CXRS measurements. It can be seen that the observed density patterns propagate in the ion-diamagnetic direction relative to $\UZExB$ at a velocity $\UZ \sim \Ubes - \UZExB \lesssim U_Z^{\rm dia,\it i}$.

Poloidal velocity profiles $\Ubes$ from the simulations with and without equilibrium flow are shown in \figref{vpol}(a, b), respectively. In case \rm{I} with adiabatic electrons, the drift velocity is small and in the ion-diamagentic direction. In simulations \rm{II} and \rm{III} (kinetic electrons, no collisions), without and with equilibrium flow respectively, the density fluctuations propagate at approximately the electron-diamagnetic drift velocity $\UZ^{\rm dia,\it e}$ with respect to $\UZExB$, which is consistent with the turbulence being driven by the trapped electrons, and in the opposite relative direction to that observed. In the cases with collisions, the trapped electron drive is reduced and consequently the fluctuations drift more towards the ion-diamagnetic direction than in the cases without collisions, more strongly in case \rm{IV} (no flow) than in case \rm{V} (with flow). Some of the cases clearly exhibit radially sheared zonal flows, which are not damped in these simulations due to the absence of collisions. As mentioned above, there are various $\mathcal{O}\lp \rhost\rp$ effects that can affect the apparent velocities, therefore, while the relative differences found between the simulation cases and observations can be considered as an indication of the underlying physics, close quantitative agreement is not expected.

\section{\label{sec:conclusions} Conclusions}

This validation exercise, comparing synthetic BES data generated from global, non-linear gyro-kinetic simulations with observations, has revealed that the simulations with the most complete physics, i.e., including kinetic electrons, sheared equilibrium flow and collisions, exhibit a degree of agreement in terms of the radial and poloidal correlation lengths, the ion heat flux and fluctuation levels (at mid-radius) and the poloidal propagation velocity of the turbulence. There are, however, notable discrepancies that suggest that the simulations have a long way to go before they can be viewed as reliably reproducing reality. Firstly, the measured correlation times are very short and always comparable to the linear timescales (drift-wave drive, parallel streaming and magnetic drift), which is consistent with strong (and probably critically balanced) turbulence, whereas the simulated turbulence exhibits much longer correlation times and appears to be only weakly non-linear. Secondly, the predicted profiles of fluctuation level and turbulent heat flux exhibit some degree of agreement with experiment only in the mid-radius region. This is partly because neo-classical physics is not included in the NEMORB simulations but also because, under conditions where the gradients are close to marginal stability, the prescribed ion-temperature gradient would have to match very precisely in order to yield the experimental heat flux. Finally, the simulations exhibit a marked shortfall in the predicted fluctuation level and ion heat flux in the peripheral region, which is likely both to be a result of the `zero-turbulence' boundary condition required to ensure numerical stability and quite plausibly some missing physics related to non-local interaction of peripheral turbulence with edge/SOL instabilities. Further work is planned to understand these result and to improve the simulations, e.g. to investigate the sensitivity of the predicted heat flux to local changes in the ion-temperature gradient, to resolve the components of the heat flux due to density and ion temperature fluctuations and to examine the possible role of parallel dynamics.

\section*{Acknowledgements}

We are grateful to I. Abel, M. Barnes, G. Colyer, S. Cowley, M. Fox, E. Highcock and F. Parra for discussions. This work was funded by the RCUK Energy Programme under grant EP/I501045 and by the European Communities under the Contract of Association between EURATOM and CCFE. To obtain further information on the data and models underlying this paper please contact PublicationsManager@ccfe.ac.uk. The views and opinions expressed herein do not necessarily reflect those of the European Commission. The gyro-kinetic simulations were performed on the HECTOR supercomputer (EPSRC Grant EP/H002081/1) and on HELIOS.

\section*{References}

%\bibliography{../ppcf_synthetic_references}

\end{document}